\documentclass[showpacs,showkeys,amssymb,aps,prl,epsfig]{revtex4}
\newcommand{\be}{\begin{equation}} \newcommand{\ee}{\end{equation}}
\newcommand{\bea}{\begin{eqnarray}}\newcommand{\eea}{\end{eqnarray}}
\usepackage{graphicx}
\begin{document}
\preprint{SINP-TNP/03-08}
\title{Near-Horizon States of Black Holes and  Calogero Models}

\author{B Basu-Mallick, Pijush K. Ghosh and Kumar S. Gupta\footnote{Talk
delivered by Kumar S. Gupta}}
\affiliation{Theory Division,\\
Saha Institute of Nuclear Physics,\\
1/AF Bidhannagar, Kolkata - 700064.}

\begin{abstract}
We find self-adjoint extensions of the rational 
Calogero model in presence of the harmonic interaction. The corresponding
eigenfunctions may describe the near-horizon quantum states of certain types
of black holes.
\end{abstract}

\maketitle

	The dynamics of particles or fields in the near-horizon region of
black holes \cite{town1,town2} is often described in terms of the
Calogero Model \cite{calo}. In particular, it has been shown that the
existence of the near-horizon conformal symmetry \cite{carlip} as well as
the logarithmic correction to the black hole entropy \cite{kaul} can be
described in terms of the self-adjoint extension \cite{reed} of the 
Calogero model in absence of the confining potential \cite{ksg,ksg1}. 
On the other hand, it has been argued that in certain
string theoretic description of black holes, the near-horizon dynamics is
governed by a many particle Calogero model in presence of the confining
potential \cite{town2}.
It is therefore of interest to find the quantum states of this
model in presence of the self-adjoint extension. These states would be
expected to encode the dynamics for such black holes. A more detailed
description of the analysis presented below can be found in Ref.
\cite{ksg2}.

The Hamiltonian of the $N$- particle rational Calogero model \cite{calo} 
is given by
\be
H = - \sum^{N}_{i=1} \frac{{\partial}^2}{\partial x_i^2} +
\sum_{i \neq j} \left [ \frac{a^2 - \frac{1}{4}}{(x_i - x_j)^2} +
\frac{\Omega^2}{16} (x_i - x_j)^2 \right ]
\label{e0}
\ee
where $a$, $\Omega$ are constants, 
$x_i$ is the coordinate of the $i^{\rm th}$ particle and
units have been chosen such that $2 m {\hbar}^{- 2} = 1$.
We are interested in finding normalizable solutions of the
eigenvalue problem $H \psi = E \psi$ in a situation where the system admits
self-adjoint extensions. We consider the above eigenvalue equation
in a sector of configuration
space corresponding to a definite ordering of particles given by
$x_1 \geq x_2 \geq
\cdots \geq x_N$. The translation-invariant
 eigenfunctions of the Hamiltonian $H$  can be written as
\be
\psi = \prod_{i <j} \left (x_i - x_j \right )^{a + \frac{1}{2}} \
\phi (r) \ P_k (x),
\label{e2}
\ee  
where $x \equiv (x_1, x_2, \dots, x_N)$,
$r^2 = \frac{1}{N} \sum_{i < j} (x_i - x_j)^2$
and $P_k (x)$ is a translation-invariant as well as  homogeneous 
polynomial of degree $k(\geq 0)$ \cite{calo}. 
The radial part of the wavefunction
satisfies the equation $\tilde{H} \phi = E \phi$ where
\be
\tilde{H} = \left [ - \frac{d^2}{dr^2} - (1 + 2 \nu )
\frac{1}{r} \frac{d}{d r} + w^2 r^2 \right ],
\ee
$w^2 = \frac{1}{8} \Omega^2 N $ and
$\nu = k + \frac{1}{2}(N - 3) + \frac{1}{2} N (N-1)(a + \frac{1}{2})$.


The Hamiltonian $\tilde{H}$ is a symmetric (Hermitian) operator on the domain
$D(\tilde{H}) \equiv \{\phi (0) = \phi^{\prime} (0) = 0,~
\phi,~ \phi^{\prime}~  {\rm absolutely~ continuous} \} $. However, when
$ -1 < \nu < 1$, $\tilde{H}$ is not self-adjoint in $D(\tilde{H})$ but
admits a one-parameter family of self-adjoint extensions \cite{reed,ksg2} 
labelled by
${\rm e}^{i z}$ where $z \in R ~({\rm mod}~ 2 \pi)$.
For $\nu \neq 0$, the spectrum is determined by the equation
\be
f(E) \equiv \frac{\Gamma \left ( \frac{ 1 - \nu }{2} - \frac{E}{4 w} \right )}
{\Gamma \left (\frac{ 1 + \nu }{2} - \frac{E}{4 w} \right ) } =
\frac{\xi_2 {\mathrm cos}(\frac{z}{2} - \eta_1)}
{\xi_1 {\mathrm cos}(\frac{z}{2} - \eta_2)},
\label{e18}
\ee 
where $\Gamma \left ( \frac{ 1 + \nu }{2} + \frac{i}{4 w} \right )
\equiv \xi_1 {\mathrm e}^{i \eta_1}$
and
$\Gamma \left ( \frac{ 1 - \nu }{2} + \frac{i}{4 w} \right )
\equiv \xi_2 {\mathrm e}^{i \eta_2}$.
This equation is plotted in the figure below.
The corresponding eigenfunctions are given by
\be
\phi (r) = B {\mathrm e}^{- \frac{w r^2}{2}}
U(d,c,w r^2),
\label{e15}  
\ee
where $c = 1+ \nu $, $d = \frac{ 1+ \nu }{2} - \frac{E}{4 w}$, 
$B$ is a constant and $U$ denotes the confluent hypergeometric function of the
second kind.

\begin{figure}
\begin{center}
\includegraphics[width=7cm]{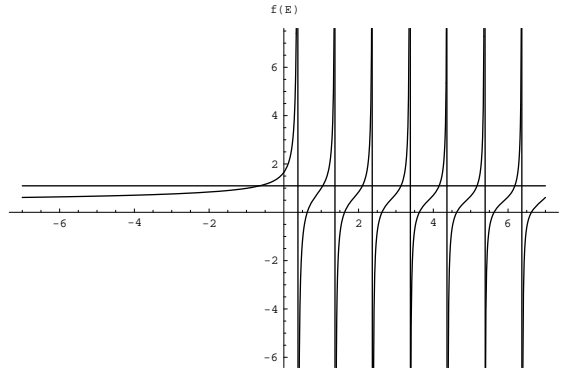}
\end{center}
\caption { \label{fig1} A plot of Eqn. (\ref{e18}) using Mathematica
with $w = 0.25$, $\nu = 0.25 $ and 
$z = -1.5$. The horizontal straight line corresponds the value of the r.h.s
of
Eqn. (\ref{e18}).} 
\end{figure}
For given values of the parameters $\nu$ and  $w$, 
the bound state energy $E$ is obtained from Eqn. (\ref{e18}) as a function of
$z$.  Different choices of $z$ thus leads to 
inequivalent quantizations of the many-body Calogero model. It may be
mentioned that the self-adjoint extensions described above exist for all
values of $N$ and for higher "angular momentum" ($k \neq 0$) sectors of the
theory as well.

The following features about the spectrum may be noted: \\
1) We have obtained the spectrum analytically when the r.h.s.
of Eqn. (\ref{e18})
is either 0 or $\infty$. When the r.h.s. of Eqn. (\ref{e18}) is $0$, we get
$E_n = 2 w ( 2 n + \nu + 1)$ where $n$ is a positive integer, corresponding
to the choice of $z = z_1 = \pi + 2 \eta_1$.
Similarly, when the r.h.s. of Eqn. (\ref{e18}) is $\infty$, 
we get $E_n  = 2 w ( 2 n - \nu + 1)$ corresponding to the value of $z$ given 
by $z = z_2= \pi + 2 \eta_2$. 
For choices of $z$ other than $z_1$ or $z_2$, the nature of the spectrum
can be understood from Figure 1, which is a plot of Eqn. (\ref{e18}) for
specific values of $\nu, z$ and $w$.
In that plot, the horizontal straight line  corresponds to the 
r.h.s of Eqn. (\ref{e18}). The energy eigenvalues are obtained from the
intersection of $f(E)$ with the horizontal straight line.
Note that the spectrum
generically consists of infinite number of positive energy solutions and at
most one negative energy solution.\\
2) Contrary to the spectrum of the rational Calogero model, 
the energy spectrum obtained from Eqn. (\ref{e18}) is not equispaced for
finite
values of $E$ and for generic values of $z$. 
This may seem surprising with the presence of $SU(1,1)$ as
the spectrum generating algebra in this system \cite{fub}, which demands 
that the eigenvalues be evenly spaced. However, when $z \neq z_1$, $z_2$,
the generator of dilatations does not in
general leave the domain of the Hamiltonian invariant
\cite{dh,ksg1}. Consequently, $SU(1,1)$ cannot be implemented as
the spectrum generating algebra except for $z = z_1$, $z_2$.\\

	It is plausible that the eigenfunctions described above would
describe the near-horizon quantum states of certain types of black holes as
described in Ref. \cite{town2}. Calculation for the corresponding density of
states and entropy would be of future interest. \\

\noindent
{\bf Acknowledgments} : The work of PKG is supported(DO No. SR/FTP/PS-06/2001) by
the SERC, DST, Govt. of India, under the Fast Track Scheme for Young
Scientists:2001-2002. 

\end{document}